\documentclass[twocolumn]{aastex63} 

\usepackage[normalem]{ulem}
\usepackage[version=4]{mhchem}
\usepackage{longtable}
\graphicspath{{./}{figures/}}
\def\farcs{\hbox{$.\!\!^{\prime\prime}$}}

\shorttitle{JWST reveals excess cool water in compact disks}

\begin{document}

\title{JWST reveals excess cool water near the snowline in compact disks, consistent with pebble drift}

\correspondingauthor{Andrea Banzatti}
\email{banzatti@txstate.edu}

\author{Andrea Banzatti}
\affil{Department of Physics, Texas State University, 749 N Comanche Street, San Marcos, TX 78666, USA}

\author{Klaus M. Pontoppidan}
\affiliation{Jet Propulsion Laboratory, California Institute of Technology, 4800 Oak Grove Drive, Pasadena, CA 91109, USA}

\author{John Carr}
\affiliation{Department of Astronomy, University of Maryland, College Park, MD 20742, USA}

\author{Evan Jellison}
\affil{Department of Physics, Texas State University, 749 N Comanche Street, San Marcos, TX 78666, USA}

\author{Ilaria Pascucci}
\affil{Department of Planetary Sciences, University of Arizona, 1629 East University Boulevard, Tucson, AZ 85721, USA}

\author{Joan Najita}
\affiliation{NSF’s NOIRLab, 950 N. Cherry Avenue, Tucson, AZ 85719, USA}

\author{Carlos E. Mu\~noz-Romero}
\author{Karin I. \"Oberg}
\affiliation{Center for Astrophysics, Harvard \& Smithsonian, 60 Garden St., Cambridge, MA 02138, USA}

\author{Anusha Kalyaan}
\affil{Department of Physics, Texas State University, 749 N Comanche Street, San Marcos, TX 78666, USA}

\author{Paola Pinilla}
\affiliation{Mullard Space Science Laboratory, University College London, Holmbury St Mary, Dorking, Surrey RH5 6NT, UK.}

\author{Sebastiaan Krijt}
\affiliation{School of Physics and Astronomy, University of Exeter, Stocker Road, Exeter EX4 4QL, UK}

\author{Feng Long}
\altaffiliation{NASA Hubble Fellowship Program Sagan Fellow}
\affiliation{Lunar and Planetary Laboratory, University of Arizona, Tucson, AZ 85721, USA}

\author{Michiel Lambrechts}
\affiliation{Center for Star and Planet Formation, GLOBE Institute, University of Copenhagen, Øster Voldgade 5-7, 1350 Copenhagen, Denmark }

\author{Giovanni Rosotti}
\affiliation{Dipartimento di Fisica, Università degli Studi di Milano, via Giovanni Celoria 16, 20133, Milano, Italy}

\author{Gregory J. Herczeg}
\affil{Kavli Institute for Astronomy and Astrophysics, Peking University, Beijing 100871, China}

\author{Colette Salyk}
\affil{Department of Physics and Astronomy, Vassar College, 124 Raymond Avenue, Poughkeepsie, NY 12604, USA}

\author{Ke Zhang}
\affil{Department of Astronomy, University of Wisconsin-Madison, Madison, WI 53706, USA}

\author{Edwin Bergin}
\affil{Department of Astronomy, University of Michigan, 1085 S. University Ave, Ann Arbor, MI 48109}

\author{Nick Ballering}
\affil{Department of Astronomy, University of Virginia, Charlottesville, VA 22904, USA}

\author{Michael R. Meyer}
\affil{Department of Astronomy, University of Michigan, 1085 S. University Ave, Ann Arbor, MI 48109}

\author{Simon Bruderer}
\affiliation{Max-Planck-Institut für extraterrestrische Physik, Gießenbachstraße 1, 85748 Garching bei München}

\author{the JDISCS collaboration}

\begin{abstract}
Previous analyses of mid-infrared water spectra from young protoplanetary disks observed with the Spitzer-IRS found an anti-correlation between water luminosity and the millimeter dust disk radius observed with ALMA. This trend was suggested to be evidence for a fundamental process of inner disk water enrichment, used to explain properties of the Solar System 40 years ago, in which icy pebbles drift inward from the outer disk and sublimate after crossing the snowline. 
Previous analyses of IRS water spectra, however, were uncertain due to the low spectral resolution that blended lines together. We present new JWST-MIRI spectra of four disks, two compact and two large with multiple radial gaps, selected to test the scenario that water vapor inside the snowline is regulated by pebble drift. The higher spectral resolving power of MIRI-MRS now yields water spectra that separate individual lines, tracing upper level energies from 900~K to 10,000~K. These spectra clearly reveal excess emission in the low-energy lines in compact disks, compared to the large disks, demonstrating an enhanced cool component with $T \approx$~170--400~K and equivalent emitting radius $R_{\rm{eq}}\approx$~1-10~au. We interpret the cool water emission as ice sublimation and vapor diffusion near the snowline, suggesting that there is indeed a higher inwards mass flux of icy pebbles in compact disks. Observation of this process opens up multiple exciting prospects to study planet formation chemistry in inner disks with JWST.
\end{abstract}

\keywords{circumstellar matter --- protoplanetary disks --- stars: pre-main sequence --- }

\section{Introduction} \label{sec: intro}

The dynamics and accretion of pebbles\footnote{Solid particles with diameters between a millimeter and a meter.} in disks are currently proposed to be fundamental for forming planets within protoplanetary disk lifetimes \citep[e.g.][]{lambrechts12,levison15}, and for determining both the mass architecture and chemical composition of planetary systems \citep[e.g.][]{ida16,bitsch19_giant,bitsch19_rocky,cridland19,lambrechts19}.
High-resolution continuum imaging with the Atacama Large Millimeter Array (ALMA) shows that protoplanetary disks with ages of a few Myr can have a wide range of sizes (10--200~au), where systems of rings and gaps are frequently observed in large disks \citep[e.g.][for recent reviews]{andrews20,bae22}. It has been suggested that the wide range in disk sizes may be due to hydrodynamical processes where large disks retain pebbles in systems of rings \citep{pinilla12}, while compact disks have experienced efficient inward ``pebble drift'' delivering solids to the inner planet forming region \citep[e.g.][]{rosotti19,appelgren20,zormpas22}. Such pebble drift should have fundamental implications for inner disk chemistry: icy pebbles that migrate inward from the outer disk will sublimate after crossing the snowline, producing a time-dependent water abundance. This systemic process was proposed to explain measured radial chemical gradients in meteorites, as well as the formation of Jupiter's core in the Solar Nebula \citep{morfill84,stevenson88,cyr98,cc06}. More recently, this process was modeled by \citet{kalyaan21} in the context of disk structures observed with ALMA, finding that the inner disk water enrichment could indeed be regulated by the presence (or absence) of gaps that retain icy pebbles in the outer disk and prevent them from entering the region of the snowline.

That inner disks ($\lesssim$ a few au) may host large columns of warm water vapor was first revealed from mid-infrared medium-resolution spectroscopy \citep{cn08,salyk08} observed from space with the Spitzer-IRS \citep{irs}. A dense forest of water emission lines, in addition to transitions and bands from OH, HCN, \ce{C2H2}, and \ce{CO2} were identified, tracing molecular budgets and chemistry in the terrestrial planet-forming region of Class II disks \citep[e.g. review by][]{pont14}. Over the past 15 years, analyses of $\approx 100$ disk spectra observed with Spitzer revealed general properties and some trends of the inner molecular reservoir, including heating by the stellar and accretion irradiation, depletion of molecules in disks with large inner dust cavities, and chemical differences as a function of stellar mass and luminosity \citep[e.g.][]{pont10,cn11,salyk11_spitz,pascucci13,walsh15,banz17,woitke18}. 
A particularly remarkable trend linking very different disk regions was recently found from the combined analysis of Spitzer spectra and ALMA continuum imaging. Expanding on earlier results from \citet{najita13}, \citet{banz20} reported an anti-correlation between the infrared water line luminosity tracing gas within a few au, and the distribution of solid pebbles at 10--200 au in disks. That is, radially compact disks show stronger water line luminosity than large disks. This correlation was tentatively interpreted in the context of pebble drift. The proposed mechanism is that pebble drift through the water snowline will enrich the inner disk oxygen abundance, and in turn the oxygen/carbon (O/C) ratio \citep[e.g.][]{bosman17,booth19,cevallos22}. A higher O/C ratio will in turn produce a higher water abundance that may be observed as a stronger water line luminosity, according to disk chemical models \citep[e.g.][]{najita11,woitke18,anderson21}. 

However, previous analyses were limited by the relatively low spectral resolving power of Spitzer-IRS (R~$\approx700$, or 450~km/s). Because of severe line blending, detailed physical properties of the emitting gas could not be retrieved to high levels of confidence. This was particularly true for water, where line blending across the forest of $\approx 1000$ emission lines led to uncertainties in column densities and abundances of 1-3 orders of magnitude \citep[e.g.][]{cn11,salyk11,banz13,james22}. While high-resolution ($R\sim 10,000-100,000$) spectroscopy from the ground can resolve individual water lines, these are limited by telluric transmission to narrow ranges of higher-excitation lines probing only an inner optically-thick hotter region \citep[500--1200~K,][]{pont10b,najita18,salyk19,banz23}. A precise analysis of water spectra across infrared wavelengths remains to date a challenge due to the complex combination of radial and vertical gradients in temperature and density, which can produce non-LTE excitation in some disk regions, and the need to observe large spectral ranges at high resolving power \citep[e.g.][]{meijerink09,bosman22,banz23}.

\begin{figure*}
\centering
\includegraphics[width=1\textwidth]{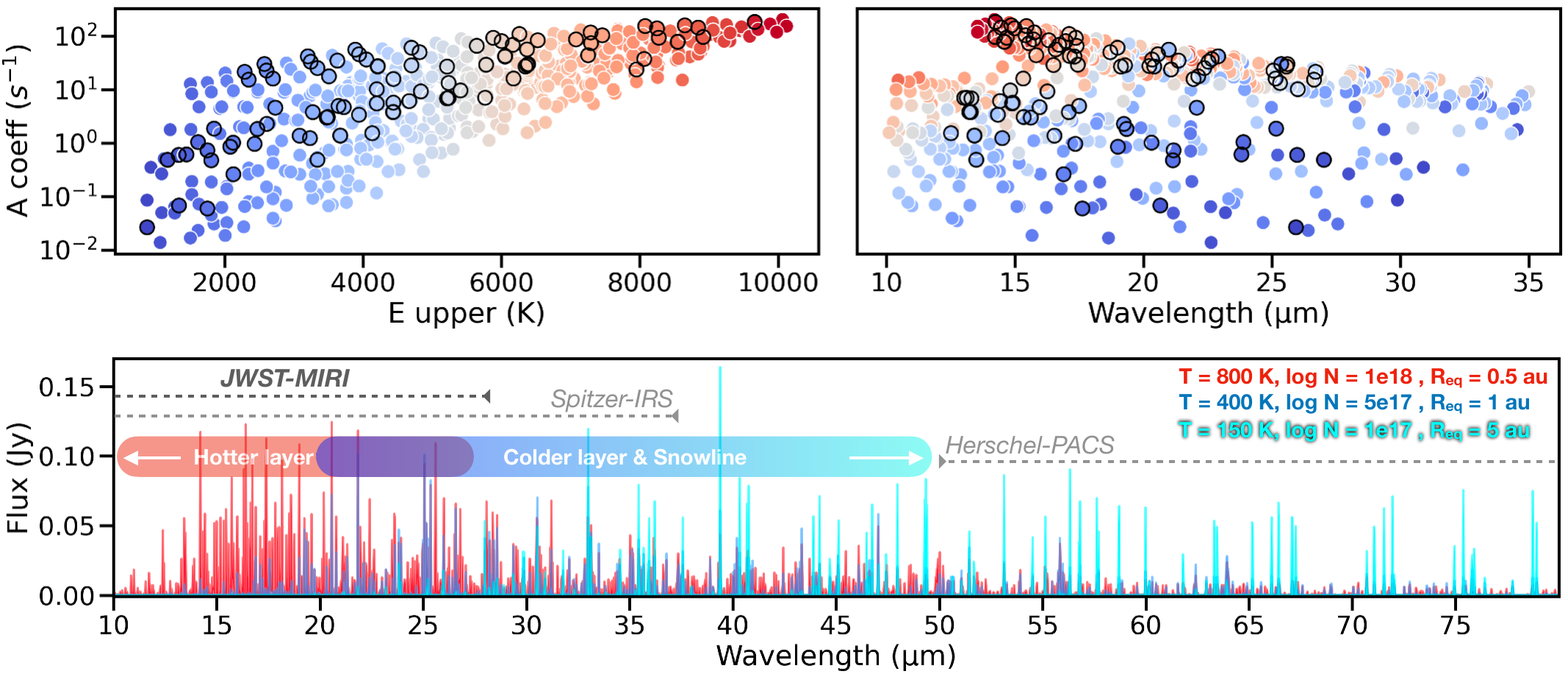} 
\caption{The distribution of upper level energies $E_u$ and Einstein-$A$ coefficients (top plots, with color-coding reflecting $E_u$ values) across infrared wavelengths naturally traces radial temperature gradients in inner disks within and across the water snowline (bottom plot, where the spectrum is scaled to a distance of 130~pc). Partly adapted from Figure 1 in \citet{blevins16} and Figure 13 in \citet{banz23}. The MIRI lines used in the analysis in Section \ref{sec: analysis} are marked with black circles.
}
\label{fig: water_radial_props}
\end{figure*}

Mitigating these challenges, the asymmetric top geometry of the water molecule leads to a very large number of rotational transitions that naturally trace radial temperature gradients, provided a wide spectral range can be observed. Water transitions from a range of upper levels (with energy $E_u \approx$ 150--10,000~K) and Einstein-$A$ coefficients ($A_{ul} \approx 0.01-100$~s$^{-1}$) extend across the near- to far-infrared wavelength range, where cooler and optically thinner transitions are generally more accessible at longer wavelengths \citep[Figure \ref{fig: water_radial_props}; see also e.g.][]{blevins16,notsu16}. Previous work combining mid- and far-infrared water spectra from Spitzer and Herschel showed that the region near the water snowline at a temperature of $\approx 150$~K dominates the observed water spectrum at $\approx$~25--180~$\mu$m \citep{kzhang13,blevins16}, whereas hotter gas within the snowline dominates the emission at shorter wavelengths (Figure \ref{fig: water_radial_props}). While the wavelength range of the Mid-Infrared Instrument (MIRI) on the James Webb Space Telescope \citep[JWST,][]{Gardner23} is therefore most focused on hotter water emission (including the ro-vibrational bands at $< 9$~$\mu$m), excess cool water emission with temperatures of $\approx $150--500\,K, just inwards of the snowline region, can still be traced at wavelengths as short as $\approx 20$~$\mu$m \citep{kzhang13,blevins16}, well within the MIRI range.

While more complex models that include global temperature gradients (150--1200~K) and non-LTE excitation are currently being explored to fit the entire range of MIRI spectra, in this paper we search for evidence for ice sublimation at the snowline in a relative sense, by following recent modeling explorations \citep{kalyaan21,kalyaan23} and addressing this question: what is the difference between water spectra in compact disks, which should be enriched in warm water due to efficient pebble drift, versus large disks with gaps, which should be relatively deficient in water due to pebble trapping at larger radii? The analysis of MIRI spectra in this paper finds evidence in support of water enrichment by pebble drift, by spectrally separating water lines that were previously blended in Spitzer-IRS spectra and revealing an excess cool water reservoir near the snowline region in compact disks. 

\begin{figure*}
\centering
\includegraphics[width=1\textwidth]{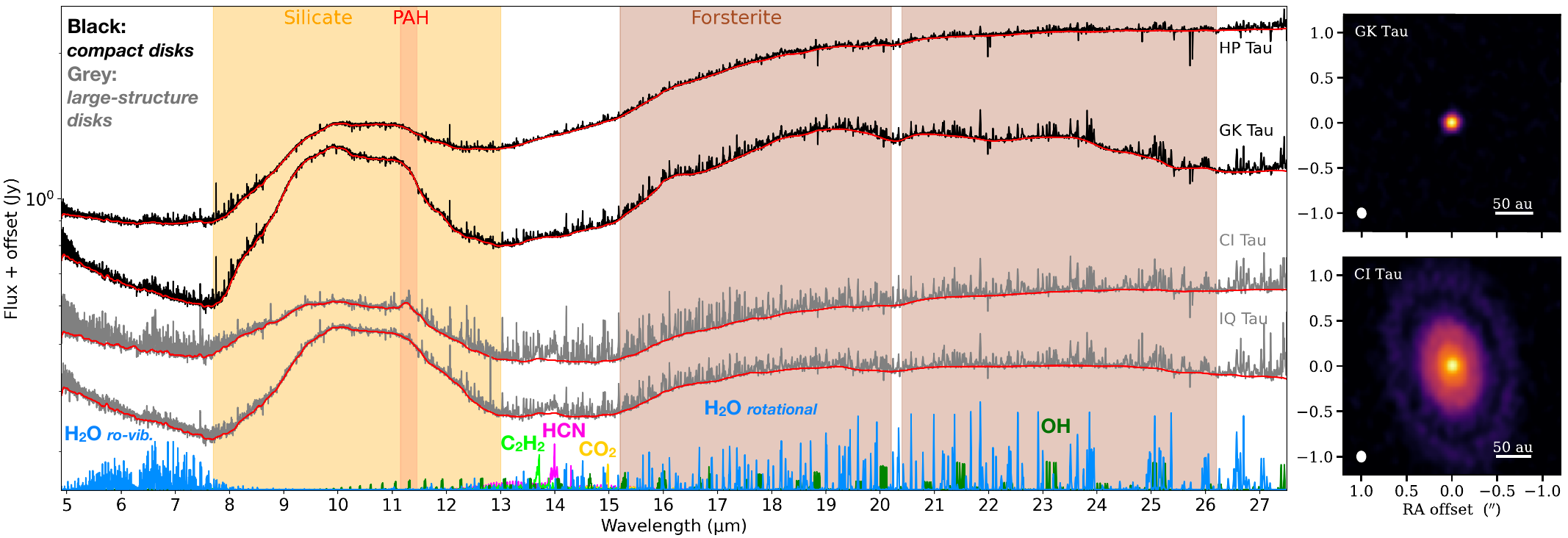} 
\caption{JWST MIRI-MRS spectra for the four disks. For clarity, the spectra are offset vertically by the following additive shifts: 0.05~Jy in CI~Tau, 0.17~Jy in GK~Tau, 0.4~Jy in HP~Tau. Illustrative models of molecular emission are shown for guidance at the bottom. Prominent dust features are approximately identified and marked with shaded regions. The estimated continuum that is subtracted before analyzing the water spectra is shown in red on each spectrum. Two ALMA images are shown to the right for reference of disk sizes and structures (the whole sample is included in Appendix \ref{app: ground data}).
}
\label{fig: JWST}
\end{figure*}

\begin{deluxetable*}{l c c c c c c c c c c c | c}
\tabletypesize{\footnotesize}
\tablewidth{0pt}
\tablecaption{\label{tab: sample} Sample properties.}
\tablehead{ Name & Dist & $T_{\rm{eff}}$ & $M_{\star}$ & $L_{\star}$ & log $L_{\rm{acc}}$ & log $M_{\rm{acc}}$ & $n_{13-26}$ & $R_{\rm{disk}}$ & Incl & CO FW10\%-50\% & $R_{\rm{CO}}$ 10-50\% & $R_{\rm{snow}}$ \\ 
 & (pc) & (K) & ($M_{\odot}$) & ($L_{\odot}$) & ($L_{\odot}$) & ($M_{\odot}$/yr) & &(au) & (deg) & (km/s) & (au) & (au) }
\tablecolumns{13}
\startdata
CITau & 160 & 4162. & 0.65 & 1.65 & -0.7 & -7.6 & -0.51 & 190.5 & 50 & 265--105 & 0.02--0.12 & 2.7 \\
GKTau & 129 & 4067. & 0.58 & 1.48 & -1.39 & -8.67 & -0.40 & 12.9 & 39 & 180--85 & 0.03--0.11 & 0.9 \\
HPTau & 177 & 4375. & 0.84 & 1.89 & -1.09 & -8.17  & 0.06 & 22.1 & 18 & 115--40 & 0.02--0.18& 1.7 \\
IQTau & 131 & 3704. & 0.42 & 0.86 & -1.40 & -8.54 & -0.74 & 109.6 & 62 & 190--150 & 0.03--0.05 & 0.9 \\
\enddata
\tablecomments{
References -- distances are from GAIA DR3 \citep{gaia_mission,gaiaDR3}, stellar and accretion properties are from \citet{hh14,donati20,manara22,gangi22}; $R_{\rm{disk}}$ and disk inclinations are from \citet{long19}; $n_{13-26}$ is measured from the MIRI spectra in this work; near-infrared CO line widths are measured at the 10\% and 50\% of the line peak from the stacked profiles shown in Appendix \ref{app: ground data}; $R_{\rm{CO}}$ is the Keplerian radius of infrared CO emission from the half line width velocities at 10\% and 50\%. $R_{\rm{snow}}$ is an expectation for the midplane snowline radius as set by accretion following Equation 2 in \citet{mulders15} as previously adopted for a general comparison across disks in \citet{banz17}.}
\end{deluxetable*}

\section{Sample \& Observations} \label{sec: data}
We present MIRI spectra of four protoplanetary disks around single stars in the Taurus star-forming region, with ages estimated to 2--3~Myr by \citet{long19} using non-magnetic evolutionary tracks from \citet{feiden16}. The disks were selected from the sample of \citet{long19} to have similar stellar and accretion luminosity, to minimize the difference in water spectra due to luminosity effects \citep{banz20}. The disks were also selected to span a wide range in measured millimeter dust disk sizes to maximize the relative difference in inner disk water abundance expected in the pebble drift scenario (Table \ref{tab: sample}). Figure \ref{fig: JWST} and Appendix \ref{app: ground data} show the ALMA images for the four disks with a spatial resolution of 0\farcs12 from \citet{long19}. The two large disks, CI~Tau and IQ~Tau, have multiple resolved gaps and rings at radial locations as small as 10~au and as large as 150~au \citep{long18,clarke18}, while the two compact disks are among the smallest known around solar-mass stars (10--20~au). The compact disks are resolved by ALMA, but do not show any sub-structures nor any millimeter emission at larger disk radii (see more in Section \ref{sec: disc}). As a measure of the pebble disk radius, $R_{\rm{disk}}$, we use the millimeter continuum radius encircling 95\% of the total integrated flux density from \citet{long19}. 

We note that HP~Tau may have a small inner dust cavity, as suggested by the relatively high infrared index $n_{13-26}$ measured in MIRI spectra instead of the Spitzer-based $n_{13-30}$ index \citep[see Appendix D in][]{banz20}. The ALMA image with 0".12 resolution does not resolve this putative cavity, even with super-resolution techniques \citep{jennings22_taurus,zhangs22}, suggesting that the cavity, if present, is smaller than $< 2$~au in width (the smallest dust structure detected with this technique in disks).

\begin{figure*}
\centering
\includegraphics[width=1\textwidth]{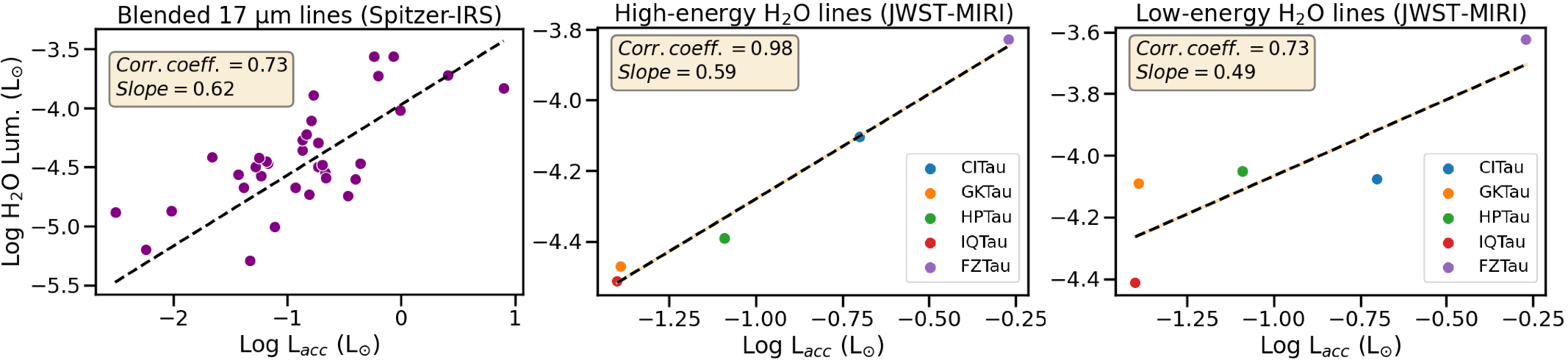} 
\caption{Correlations between water line luminosities and accretion luminosity. \textit{Left:} using blended water lines at 17.1--17.4~$\mu$m from Spitzer-IRS \citep[data and regression adopted from][]{banz20}. \textit{Middle:} using high-energy (6000--8000~K) water lines from JWST-MIRI from this work and Paper I; \textit{right:} using low-energy (900--3000~K) water lines (see Figure \ref{fig: water_radial_props}).}
\label{fig: Lacc_correlations}
\end{figure*}

The disks were observed over the full wavelength coverage of 4.9--28\,$\mu$m with the Medium Resolution Spectrometer \citep[MRS,][]{jwst-mrs} mode on MIRI \citep[][]{miri,miri2} as part of program GO-1640 in Cycle 1. Figure \ref{fig: JWST} shows the four MIRI-MRS spectra obtained on February 27-28, 2023 with deep integrations ($\approx 1000$~s in GK~Tau and HP~Tau and $\approx 1800$~s in CI~Tau and IQ~Tau) and reduced as described in \citet{pont23} (henceforth Paper I from the JDISCS collaboration). Two standard calibrators (asteroids) from program GO-1549 were used to maximize the quality of fringe removal, the spectral response function, and flux calibration across the four MIRI channels (Paper I). Target acquisition was used to ensure pointing precision and match across different targets and the asteroids to sub-pixel level. An improved wavelength calibration was applied from cross-correlating 200 molecular lines across MIRI wavelengths, improving the precision from $\approx 90$~km/s down to better than 5~km/s. The spectra were continuum-subtracted using an iterative median filter with a box of $\approx 100$ pixels and a smoothing step using a 2nd-order Savitzky-Golay filter (Paper I). The MIRI spectra and continua are shown in Figure \ref{fig: JWST}.

As the MIRI resolving power of $R=1500-3500$ ($90-200~\rm km\,s^{-1}$) cannot resolve the gas kinematics \citep{miri_res}, in Appendix \ref{app: ground data} we report for reference high-resolution ($R\approx 6-9\times 10^{4}$, or $3-5\,\rm km\,s^{-1}$) line profiles for the CO fundamental band as observed from the ground. CI~Tau and GK~Tau have comparable CO line profiles, suggesting that their molecular gas radial distribution in the inner disk should be similar. IQ~Tau shows more compact emission, possibly due to the higher disk inclination \citep{banz22}, while HP~Tau may have a small inner dust cavity (see above). Therefore, while we study all four disks, our analysis below will particularly focus on the comparison between CI~Tau (large disk with gaps) and GK~Tau (compact disk).

\begin{figure*}
\centering
\includegraphics[width=1\textwidth]{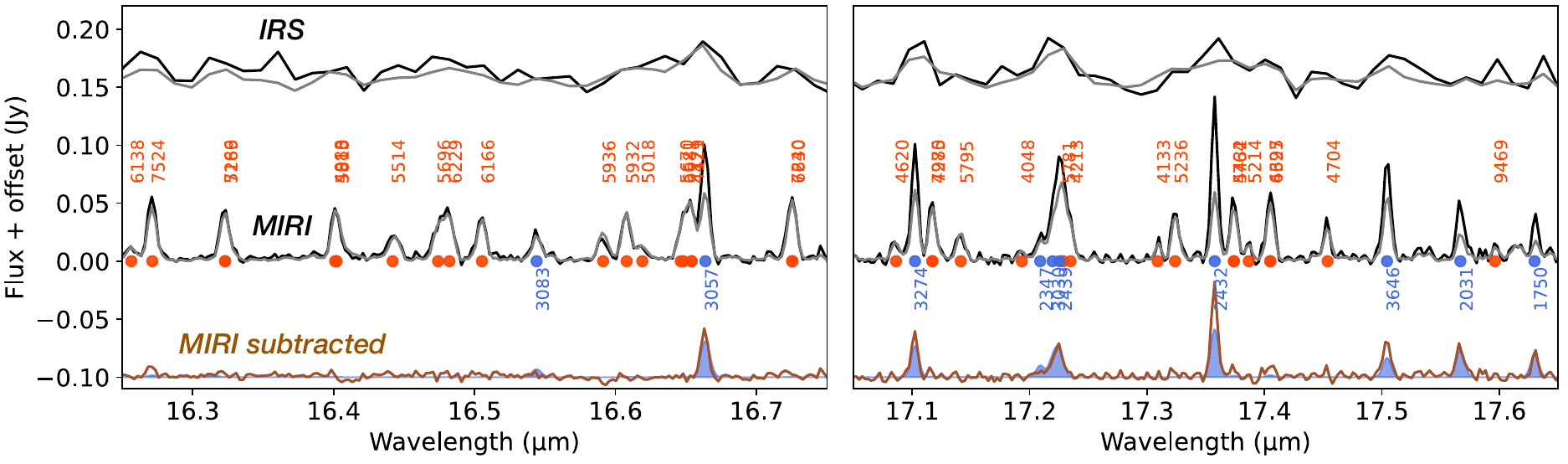} 
\caption{Sections of the continuum-subtracted infrared water spectra of the compact disk GK~Tau (in black) and the large disk CI~Tau (in grey), at the top from Spitzer-IRS (offset by 0.15 Jy) and at the center from JWST-MIRI. A broader range of the spectrum is shown in Figure \ref{fig: residual_spitzer} in Appendix \ref{app: cool component}. The CI~Tau spectrum is scaled to the distance of GK~Tau and corrected for the different accretion luminosity (Section \ref{sec: analysis}). The subtraction of the two spectra is shown in brown at the bottom, with the single-temperature LTE model (blue area) from Figure \ref{fig: cool_residuals} and Table \ref{tab: slab_results}. Vertical labels show the upper level energy $E_u$ in K, separated into higher-energy ($> 4000$~K) in red and lower-energy ($< 4000$~K) in blue.}
\label{fig: subtraction_zoomed}
\end{figure*}

\section{Analysis \& Results} \label{sec: analysis}
As summarized in Section \ref{sec: intro}, cool water emission from the region near the snowline should dominate the emission observed in lower-energy levels at longer infrared wavelengths. However, different temperatures contribute different fractions of line flux across a wide range of wavelengths depending on the upper level energy of a specific transition (Figure \ref{fig: water_radial_props}). While single-temperature LTE models have been used to successfully reproduce limited portions of water spectra observed with MIRI \citep[][and Appendix \ref{app: hot component}]{grant23}, modeling the entire MIRI wavelength range should generally require to include multiple temperature components, or a radial temperature gradient \citep{blevins16,liu19,banz23}. Further, non-LTE excitation is likely important for the higher energy lines with $E_u \gtrsim 4500$~K \citep{meijerink09,banz12}. Since such modeling is computationally expensive and will suffer from some degree of degeneracy, to address the specific question of this work we employ an empirical method to separate representative temperature components by using the observed water spectra themselves as templates for comparison between small vs large disks, as follows. 

Spitzer-IRS water spectra from disks have been found to be very similar at 12--16~$\mu$m as observed with Spitzer-IRS, suggesting that a different emitting area (i.e. a simple scaling factor that does not change line ratios in this range) could account for most of the difference in water luminosity between different sources \citep{salyk11_spitz,cn11}. Previous work also found that the dominant correlation with water luminosities (for disks with detected water emission) is with accretion luminosity \citep{banz20}, consistent with UV irradiation from stellar accretion playing an important role in the heating and excitation of water by driving the size of the emitting area \citep[e.g.][]{walsh15,najita17,woitke18,bosman22}.  Comparisons of water spectra across different disks therefore require normalization of the total water line luminosity to remove the primary effect of a luminosity-dependent emitting area from disk to disk. Indeed, the anti-correlation between water emission and the millimeter dust disk radius was found after normalizing the measured water luminosity with the stellar accretion luminosity \citep[Figure 6 in][]{banz20}. 

To test the correlations previously found with water luminosity, we measure the luminosities of $\approx 100$ individual water rotational lines at 12--27~$\mu$m within the new MIRI spectra. The lines were selected to be spectrally separated from other water lines, as well as from lines from other molecular species known to emit at these wavelengths (OH, HCN, \ce{C2H2}, \ce{CO2}, \ce{H2}, \ce{HI}, \ce{NeII}). The selected water lines span $E_u$ between 900 and 10,000~K (black circles in Figure \ref{fig: water_radial_props}). By testing correlations as a function of $E_u$, we find that the primary correlation with $L_{\rm{acc}}$ consistently increases with $E_u$. That is, higher-energy lines correlate more strongly with $L_{\rm{acc}}$. This is concisely demonstrated in Figure \ref{fig: Lacc_correlations} by considering lines into two bins: higher-energy lines with $E_u =$~6000--8000~K (with a correlation coefficient of 0.98) and lower-energy lines with $E_u =$~900--3000~K (with a correlation coefficient of 0.73). To increase the leverage for characterizing the correlation we include also the high-accretor FZ~Tau from Paper I. The correlation in the higher-energy lines is not significantly affected by the removal of the FZ~Tau data point (Figure \ref{fig: Lacc_correlations}). The larger correlation coefficient with larger $E_u$ is consistent with the stellar and accretion irradiation setting the emitting area of primarily the hotter, optically-thick, inner disk layer (as shown in Appendix \ref{app: hot component}), whereas cooler gas at larger radii may include additional effects from processes linked to ice sublimation near the snowline (as discussed in Section \ref{sec: disc}). The linear regression of the high-energy line luminosity with accretion luminosity has a slope of $\approx 0.6$, which confirms the value previously found using blended lines at 17.1--17.4~$\mu$m from Spitzer-IRS in \citet{banz20}. The de-blended lines from a wider range of energy in the MIRI spectra now indicate that the correlation was mostly driven by the higher-energy lines, which indeed dominate the 17.1--17.4~$\mu$m water lines (Figure \ref{fig: subtraction_zoomed}). 
\begin{figure*}
\centering
\includegraphics[width=1\textwidth]{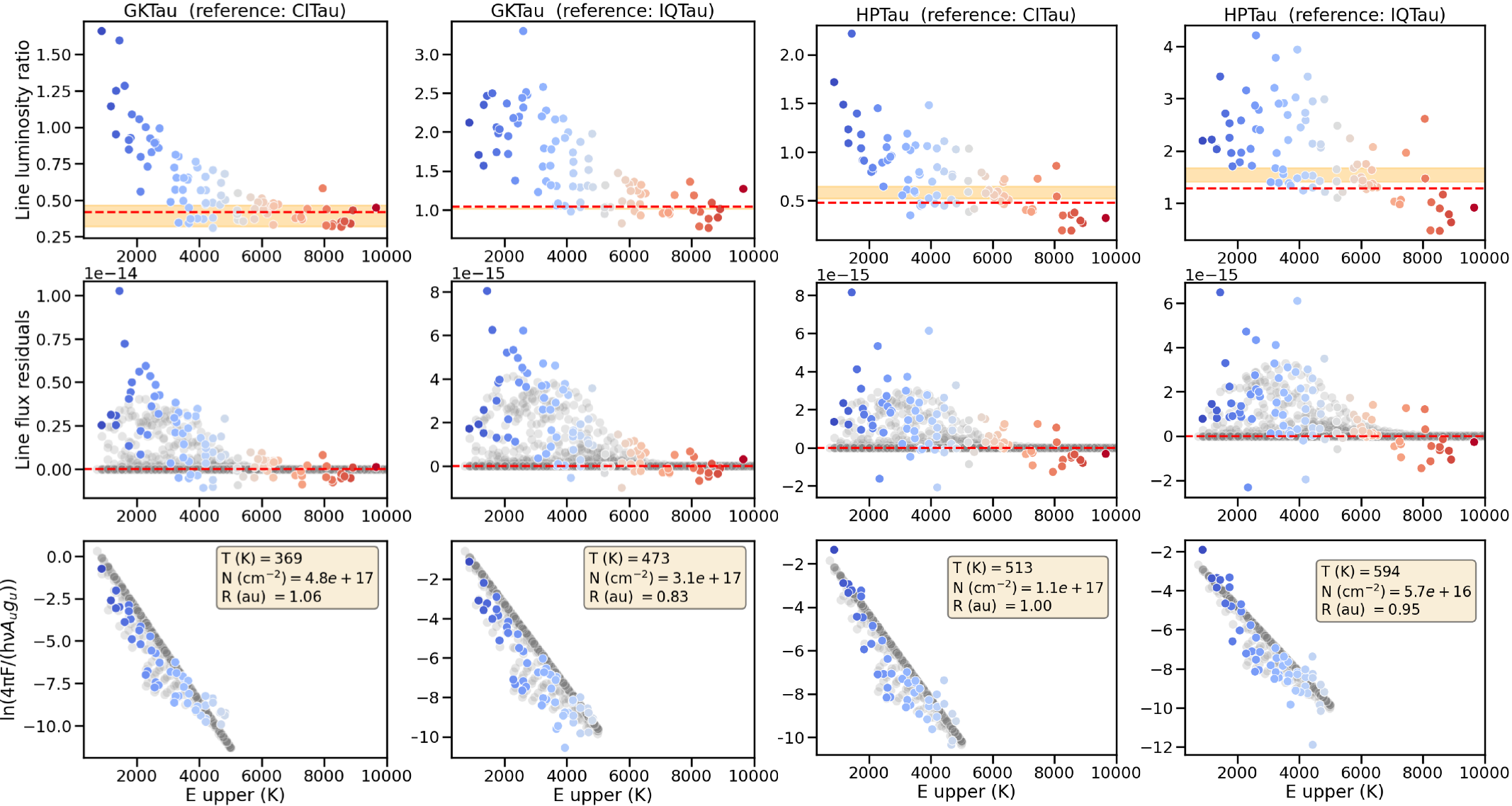} 
\caption{Overview of excess cool components in compact disks by using water spectra in large disks as reference templates for the hotter inner component, as explained in Section \ref{sec: analysis}. The orange area in the top plots is the high-energy luminosity ratio expected from using the correlation with $L_{\rm{acc}}$ shown in Figure \ref{fig: Lacc_correlations} (3-$\sigma$ region). The red-dashed line is the median value of the line luminosity ratio at $E_u \sim$~6000--9000~K. The datapoint color gradient reflects the $E_u$ value (x axis) of each water line. A single-temperature LTE fit to line fluxes measured in the excess cool component is shown in grey in the middle (line flux residuals) and bottom (population diagram) plots, with parameters shown in the box. }
\label{fig: cool_residuals}
\end{figure*}

In Figure \ref{fig: subtraction_zoomed}, the continuum-subtracted MIRI spectra of GK~Tau (compact disk) and CI~Tau (large disk with multiple gaps) are compared after scaling CI~Tau to the same distance of GK~Tau, and then applying to CI~Tau a scaling factor of 0.42 from the luminosity ratio of the high-energy water lines to remove the line emitting area dependence on accretion luminosity according to the correlation in Figure \ref{fig: Lacc_correlations}. The two spectra are then subtracted to obtain the spectral difference in water excitation, revealing that while the relative excitation of high-energy lines is extremely similar in the two disks (after the common scaling factor from the luminosity-dependent emitting area is accounted for), the low energy lines consistently show some flux excess in the compact disk of GK~Tau. The figure also demonstrates that the lower resolution of previous Spitzer spectra did not allow for an unambiguous detection of the cool water excess, although, in hindsight, a slight excess in correspondence of the lower-energy lines may now be identified in the Spitzer-IRS spectrum of GK~Tau (Figure \ref{fig: subtraction_zoomed}).

\begin{figure*}
\centering
\includegraphics[width=1\textwidth]{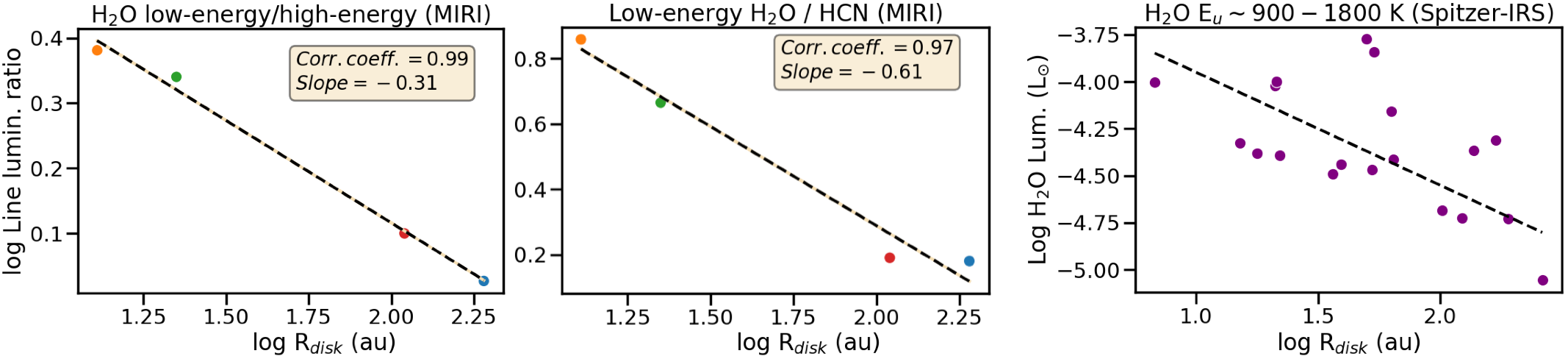} 
\caption{Correlations with dust disk radius $R_{\rm{disk}}$ for the luminosity ratio between resolved water low-energy and high-energy lines in MIRI (left), resolved water low-energy lines and the 14~$\mu$m HCN band in MIRI (center), and blended water low-energy lines in Spitzer near 30~$\mu$m \citep[taken from Figure 9 in][excluding disks with large inner dust cavities]{banz23}. }
\label{fig: Rdisk_correlations}
\end{figure*}

The procedure of scaling and subtracting a template spectrum of a large disk from that of a compact disk is applied systematically to the sample included in this work as visualized in Figure \ref{fig: cool_residuals}. In each column, line luminosities from a compact disk (GK~Tau or HP~Tau) are divided by those from a large disk (CI~Tau or IQ~Tau) to obtain the empirical scaling factor between the higher-energy lines tracing the inner hotter emitting region (from the median value of line luminosities between $E_u = $~6000--9000~K). The expected scaling factor from applying the linear correlation of the high-energy line luminosities (Figure \ref{fig: Lacc_correlations}) is shown in orange for reference. The high-energy spectral template is then subtracted, and the line flux of the residuals is fitted with a single-temperature LTE model \citep{banz12,salyk11_spitz,salyk20}. 
The excess cool component in the two compact disks is reproduced by an LTE model with rotational temperature $T\approx $~400--500~K, column density $N \approx 1-5 \times 10^{17}$~cm$^{-2}$, and equivalent emitting radius $R_{\rm{eq}}$~$\approx 0.8-1$~au. 
The residuals and best-fit solutions are visualized in the rotation diagram in the bottom part of Figure \ref{fig: cool_residuals}. 

We note some dependency of the choice of template spectrum on the measured parameters of the cool excess: $T$ is higher while $N$ and $R_{\rm{eq}}$ are lower when using IQ~Tau as the template instead of CI~Tau (second and fourth columns in Figure \ref{fig: cool_residuals}). We interpret this as due to a subtle difference in excitation for the higher-energy lines in the two large disks combined with the intrinsic degeneracies of single-temperature slab model fits \citep{salyk11_spitz,cn11}. Further, the temperature of the cool excess in HP~Tau is $\approx 100$~K higher than in GK~Tau, possibly as a consequence of differences in the inner disk physical structures, as suggested by the higher infrared index $n_{13-26}$ in HP~Tau.

\begin{deluxetable}{l c c c c c c}
\tabletypesize{\footnotesize}
\tablewidth{0pt}
\tablecaption{\label{tab: slab_results} LTE model fits to rotational water lines.}
\tablehead{\colhead{Object} & \multicolumn{3}{|c|}{H$_{2}$O high energy} & \multicolumn{3}{c}{H$_{2}$O low-energy excess} 
\\ 
 & $R_{\rm{eq}}$ & $T$ & log~$N$ & $R_{\rm{eq}}$ & $T$ & log~$N$  
\\
 & (au) & (K) & (cm$^{-2}$) & (au) & (K) & (cm$^{-2}$)   }
\tablecolumns{7}
\startdata
CI~Tau & 0.65 & 840 & 1.0e18 & -- & -- & --   \\
IQ~Tau & 0.38 & 850 & 1.1e18 & -- & -- & --   \\
HP~Tau & 0.46 & 820 & 1.1e18 & 1.00 & 510 & 1.1e17   \\
GK~Tau & 0.45 & 830 & 7.1e17 & 1.06 & 370 & 4.8e17   \\
GK~Tau$^a$ & -- & -- & -- & 0.87 & 420 & 3.5e17   \\
GK~Tau$^a$ & -- & -- & -- & 8.6 & 170 & 3.1e16   \\
\enddata
\tablecomments{See Section \ref{sec: analysis} for details on the separation of components. For the high-energy component, we report fit results to the 12--16~$\mu$m range from Appendix \ref{app: hot component}. For the cool excess component, we report results obtained by using the spectrum of CI~Tau as template for the inner hot component (see Figure \ref{fig: cool_residuals}). While IQ~Tau shows some cool excess in comparison to CI~Tau (Figure \ref{fig: Rdisk_correlations}), it is too weak for the fit to converge. $^a$Results from the two-temperature fit to MIRI and IRS residuals reported in Appendix \ref{app: cool component}.}
\end{deluxetable}

Lastly, we consider the relations between the properties of the excess cool component and the ALMA dust disk radius in Figure \ref{fig: Rdisk_correlations}. The first panel shows the ratio between the line luminosities of the low-energy and high-energy water transitions from MIRI. With the four targets analyzed in this work the correlation coefficient is very high (0.99); this should be more globally characterized in the future using larger samples. The middle panel shows the ratio between the same low-energy water lines, but now using the luminosity of the 14\,$\mu$m HCN band at the denominator, after subtraction of a hot-component water model (Appendix \ref{app: hot component}). This is shown for reference to previous works that measured the \ce{H2O}/HCN ratio from Spitzer spectra \citep{najita13,banz20}. 
As proposed in \citet{banz20}, the ratio with HCN may work as a normalization factor for \ce{H2O} emission, but a change in the O/C elemental ratio may also influence the relative luminosity of \ce{H2O} versus organic molecules, which will be analyzed in more detail in future work. In particular, carbon grain destruction near the soot line at $\approx 500$~K \citep{vanthoff20,li21,tabone23} could result in a different O/C ratio between the regions traced by the hotter ($\approx 800$~K) vs colder ($\approx 400$~K) water components analyzed in this work.

The third panel shows a previous correlation reported in \citet{banz23} using low-energy water lines near 30.7~$\mu$m from Spitzer spectra. These lines are blends of transitions with $E_u = 900-1800$~K, are well-separated from contamination by known OH lines, and were found to have a stronger anti-correlation with the disk radius than the higher-energy lines at shorter wavelengths \citep{banz23}. A re-analysis of archival Spitzer-IRS spectra of GK~Tau and CI~Tau following the procedure developed above for the MIRI spectra now shows that excess water emission from a temperature consistent with ice sublimation ($T \approx 170$~K) with a large emitting area ($R_{\rm{eq}}$~$\approx 9$~au) dominates the \ce{H2O} lines at $> 30$~$\mu$m in the compact disk of GK~Tau (Appendix \ref{app: cool component}), with observable flux in MIRI in two low-energy lines at 23.8--23.9~$\mu$m as expected from previous works (Section \ref{sec: intro} and Figure \ref{fig: water_radial_props}). We can therefore conclude with confidence that there is strong evidence from all the available data, including JWST-MIRI and Spitzer-IRS, for excess cool water vapor in the compact disks as compared to the large disks in this sample.

\section{Discussion} \label{sec: disc}
In this work we have analyzed MIRI-MRS spectra of water rotational emission from four protoplanetary disks, two compact ($R_{\rm disk} \approx 10-20$~au) and two large ($R_{\rm disk} \approx 100-150$~au) with multiple dust gaps observed with ALMA, to test previous evidence for enrichment of inner disk water vapor by ice sublimation at the snowline in drift-dominated disks \citep[][]{banz20}. The analysis shows that the observed higher line luminosity in the compact disks is due to excess flux in the low-energy lines with $E_u < 4000$~K (Figure \ref{fig: subtraction_zoomed}). While all disks share a similar component in the higher-energy lines of optically-thick vater vapor with $T \approx 800$~K and $N \approx 1 \times 10^{18}$~cm$^{-2}$ (see Appendix \ref{app: hot component}), the compact disks show excess emission consistent with $T \approx 400$~K and $N \approx 1 \times 10^{17}$~cm$^{-2}$ (Section \ref{sec: analysis}), which extends down to at least $T \approx 170$~K when a wider range of lower-energy lines with $E_u < 2000$~K are included, from wavelengths longer than those covered by MIRI-MRS (see Appendix \ref{app: cool component} using Spitzer-IRS). 
The presence of multiple temperature components probably approximates a temperature gradient -- and related radial abundance structure -- in the inner disk surface. The existence of a temperature gradient, and in some cases distinct gas reservoirs, is a general prediction of models that include water chemistry and/or transport in planet-forming disks (see references in Section \ref{sec: intro}). This is consistent with general gradients previously retrieved using spatially- and spectrally-unresolved Spitzer and/or Herschel spectra \citep{blevins16,liu19}. Within the hotter water emission component, a range of emitting regions and temperatures have been proposed based on spectrally-resolved line widths that decrease between $E_u \approx 10,000$~K at least down to $E_u \approx 4000$~K, the energy range that is currently accessible to high-resolution mid-infrared spectroscopy from the ground \citep{banz23}. 

\begin{figure*}[ht!]
\centering
\includegraphics[width=1\textwidth]{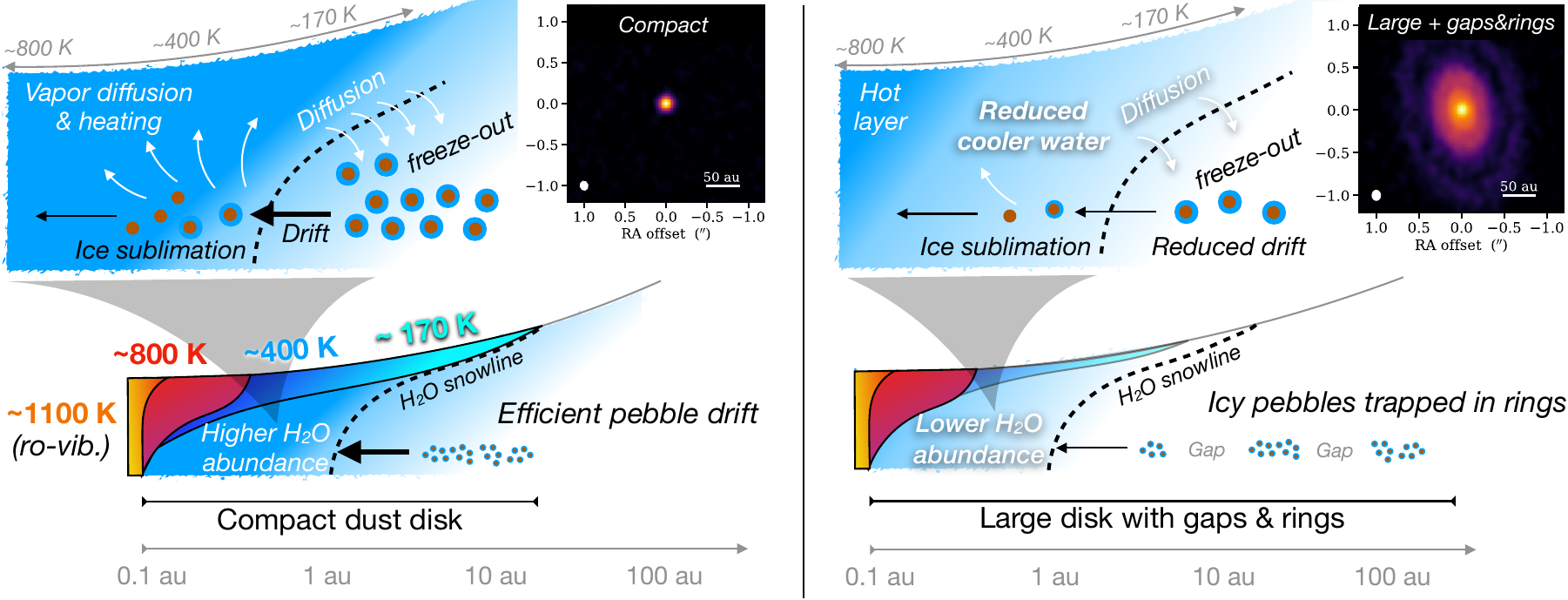} 
\caption{Illustration of the interpretation of the results in the context of the long-established paradigm of inner water enrichment by pebble drift \citep{cyr98,cc06}. The insets show the main processes of pebble drift, ice sublimation, and vapor diffusion as included in recent model explorations by \citet{kalyaan21,kalyaan23} in the context of how ice delivery and inner disk water enrichment may be regulated by dust gaps observed with ALMA. }
\label{fig: cartoon}
\end{figure*}

From the MIRI spectra, the slab model fits reported in this work indicate that the hotter component emits from a more compact region ($R_{\rm{eq}}$~$\approx 0.5$~au), likely from smaller disk radii, while the excess cooler component traces a more radially extended region ($R_{\rm{eq}}$~$\gtrsim 1$~au). While large disks may still have weak emission from the cooler region, this work clearly demonstrates that the two compact disks have a significant excess reservoir of cool water vapor, consistent with a larger emitting area. We visualize the proposed approximate radial and vertical location of these different water components in Figure \ref{fig: cartoon}. We remark that the moderate resolution of MIRI does not provide direct information on the emitting disk radii for the two water components, only equivalent emitting areas (Section \ref{sec: analysis}); a direct characterization of spatial regions needs higher spatial or spectral resolution from future instrumentation.

The presence of a hot, inner water layer may be a common feature of gas-rich inner disks around $\sim$solar-mass young stars as found in previous observations with Spitzer-IRS \citep{salyk11_spitz,cn11}. Indeed, it is found to have very similar temperature and column density, and only a different emitting area, in all four disks in this work regardless of their outer dust disk size (Appendix \ref{app: hot component}). The temperature of $\approx 800$~K may represent an average of a temperature gradient extending at least up to $\approx 1200$~K in a more compact inner region at the dust sublimation radius, as suggested by analysis of hot ro-vibrational water lines \citep{banz23}. This warm-hot layer is commonly predicted by thermo-chemical disk models where water vapor forms efficiently through gas-phase chemistry \citep[which is otherwise very inefficient at low temperatures, $T < 300$~K, e.g.][]{glassgold09}. These model predictions indeed limit the observed infrared water emission to a surface layer with $T \approx 400-1000$~K and column of $N \approx 10^{18-19}$~cm$^{-2}$ \citep[e.g.][]{najita17,woitke18,anderson21,bosman22}, in dynamically static scenarios that exclude the effects of radial drift of icy pebbles. The ubiquitous presence of the hot component therefore seems consistent with static chemical models, and does not necessarily require an enrichment from icy pebbles drifting from the outer disk. The presence of excess cool water emission in the compact disks, instead, provides an interesting new element to the picture. 

What gas reservoir is the excess cool water emission tracing? The large difference in the best-fit emitting areas indicates that the vertical overlap between the hotter and cooler emitting regions is necessarily limited. The cooler water component must be tracing larger radii rather than just a deeper layer. In fact, while a vertical stack of line emission from a given energy level can produce saturation that may hide cooler emission from closer to the disk midplane, radially distributed line emission will add up in flux without saturation when different temperature components do not share lines of sight. The lower-energy lines also globally have lower Einstein-$A$ coefficients (0.1--10~s$^{-1}$, compared to 10--100~s$^{-1}$ in the high-energy lines, Figure \ref{fig: water_radial_props}), and therefore naturally trace cooler gas extending to larger radii near and across the snowline \citep{kzhang13,blevins16}. Radially extended emission in the low-energy lines at temperatures $T \approx$~170--400~K therefore most likely traces a region very close to the ice sublimation front, as ice in disks is indeed expected to sublimate at temperatures $T \approx$~120--180~K \citep[e.g.][]{pollack94,sasselov00,lodders03} and vapor is heated up as it diffuses toward hotter disk layers. 

Why is this cooler water emission enhanced in compact disks? \citet{kalyaan21,kalyaan23} recently presented a series of models investigating the process of inner disk water enrichment from drifting icy solids \citep{cyr98,cc06} in the context of dust gaps and rings revealed by ALMA imaging \citep{andrews20}. The models by Kalyaan et al. included both gas and dust evolution, with radially drifting icy solids that progressively sublimate after entering the snowline region, releasing vapor that diffuses inward/upward to enrich the inner disk vapor, and outward across the snowline to freeze-out again on icy grains (see schematic in Figure \ref{fig: cartoon}). They found that inner disk water enrichment is suppressed when a sufficiently deep gap providing an efficient dust trap is present in the disk, particularly when the gap is located at smaller radii close to the snowline, i.e. when it prevents a larger fraction of the outer ice reservoir from reaching the snowline. Disk viscosity is also expected to play a role in regulating pebble drift and the inner disk water enrichment, with lower disk viscosities increasing the inner disk water abundance \citep{schneider21,kalyaan23}. As the outer dust disk radius shrinks under the effect of inward pebble migration in drift-dominated disks \citep[e.g.][]{rosotti19,appelgren20,zormpas22}, an anti-correlation between inner water enrichment and outer disk radius could indeed be produced as a consequence of icy pebble trapping beyond disk gaps \citep[Figure 10 in][]{kalyaan21}. Models also showed that in absence of drift a vertical ``cold finger effect'' could quickly deplete the surface layer above the midplane snowline, both for water and for CO \citep{meijerink09,krijt18}, whereas that region would be replenished through ice sublimation and diffusion whenever radial drift is active. 

While models propose that it should be the innermost gap to regulate the inner disk water enrichment through pebble drift \citep{kalyaan21}, the observed dust disk size is instead set by the location of the outermost gap that retains enough pebbles to be detected with ALMA. The link between the inner disk cool water excess and the outer dust disk size could therefore not necessarily be ubiquitous, and depend instead on the number, depth, and location of gaps in disks \citep{kalyaan23}. It is worth noting that large disks may have multiple pebble-stopping gaps across a range of disk radii, and this is surely the case of the two large disks included in this work. IQ~Tau has at least two gaps located at $\approx$~40 and 60~au, and CI~Tau has three gaps located at $\approx$~14, 50, 120\,au \citep{long18}, and possibly an additional gap at 5~au \citep{jennings22_taurus}. Moreover, CI~Tau has higher-contrast gaps even in comparison to IQ~Tau, which are likely more efficient in trapping pebbles \citep[e.g.][]{pinilla12}, potentially explaining why this is the disk with the lowest cool water excess among the large disks in this work (Figure \ref{fig: Rdisk_correlations}). 

While compact disks may have narrow gaps at $< 30$~au that remain unresolved by ALMA, none have been reported for the two compact disks in our sample, even by using super-resolution techniques \citep{jennings22,jennings22_taurus,zhangs22}. 
Further, shallow gaps in inner disks may be very leaky and effectively produce little or no reduction in inner water enrichment \citep{kalyaan23}. Future high-resolution dust continuum imaging is needed to better understand the presence of the innermost disk dust gaps and their efficiency in trapping pebbles. While this objective is at the spatial resolution limits of ALMA, it may be possible with a future facility, such as the next generation Very Large Array \citep[ngVLA, e.g.][]{andrews_ngvla}. A reduced excess in the cool water reservoir may provide high-priority disk targets for future high-resolution observations to spatially resolve gaps near the snowline at 1--5~au.

In summary, pebble drift and trapping provides a fundamental, natural process for a large-scale link between inner and outer disk regions that may explain the cool water excess revealed by MIRI in the compact disks analyzed in this work. With additional high-resolution data for larger samples, future work will establish how general this effect may be and further establish if pebble drift is indeed the cause of the cool water excess. The specific targets included in this work were observed to test the pebble drift hypothesis by selecting similar-luminosity stars with very different disk sizes (Section \ref{sec: data}), with the two large disks having multiple gaps across disk radii reducing the inward pebble drift \citep{pinilla12}. Pebbles are, however, sufficiently large to settle in the disk midplane and cannot be directly observed in the innermost optically-thick disk regions, nor can the midplane water vapor mass be directly measured; indeed, the observed surface water would need to be mixed up from the midplane, likely by diffusion and turbulence. While the new observational results from MIRI add evidence that is consistent with water delivery by pebble drift, we consider in the following other scenarios that may offer additional or alternate processes that regulate the inner disk water abundance.

\paragraph{Inner disk dust cavities:} The presence of a large inner dust cavity (gaps that extend to the inner dust radius) has been found to likely deplete the abundance of multiple molecules, including water \citep{salyk11_spitz,salyk19,banz17,banz20,banz22}. Yet, a small cavity may not have the same effects. All four disks in this paper show CO rovibrational emission tracing hot gas down to within 0.1~au (Appendix \ref{app: ground data}). This demonstrates that there is molecular gas at small radii, without evidence for a gas cavity. The only disk for which the mid-infrared spectral index suggests the presence of an inner dust disk cavity, HP~Tau, is a compact disk where the cool water excess is detected.  
If HP~Tau indeed has a $< 2$~au gas-rich dust cavity (to be undetected in the ALMA continuum image) and a mid-plane snowline of $\approx 1$~au (Table \ref{tab: sample}), icy pebbles may still drift through the snowline and provide the cool water excess. A potential alternate explanation for a reduced reservoir of cool water vapor in CI~Tau, instead, could be if an inner gap at 5~au \citep{jennings22} is in the region of the surface snowline, and it might be depleting the cool water layer (e.g. by dissociation of \ce{H2O}) in addition to reducing pebble drift.

\paragraph{Inner disk gas/dust layers:} Could the excess cool water emission be the consequence of a different gas/dust physical structure in compact disks, specifically different gas-to-dust ratios or gas densities? If the compact disks are less dense in their inner region, we may be detecting cooler water deeper towards the disk mid-plane. While a deeper layer would not be visible below the optically thick hotter inner water, it might become visible in the less optically thick cooler component. The observability of these layers may also be linked to the relative vertical distribution of gas and dust. In a static disk model, infrared molecular spectra are found to be strongly affected by the gas-to-dust ratio in the emitting layer \citep[e.g.][]{woitke18}. In the pebble drift scenario, the dust-to-gas ratio would change in time as a function of the inward drifting pebbles in parallel to the water enrichment \citep{kalyaan21}. While molecular line fluxes are generally expected to increase with the decrease of small dust grains in the disk surface \citep{greenwood19}, the specific effects of a time-variable dust-to-gas ratio on the observed water spectrum (including the layer depth for different temperature components) will require dedicated future work to be quantified.

\paragraph{Disk surface accretion:} Surface accretion flows may offer an alternate way to transport water to inner disk atmospheres. In contrast to pebble accretion, in which icy material close to the disk midplane flows inward rapidly relative to the gas, small icy grains near the disk surface can flow inward as part of the general accretion flow that transports gas and dust through the disk toward the star. While surface flows deliver water in the form of well-mixed small icy grains, this process will not enrich water over the primordial baseline, but it may replenish observable surface water lost by freeze-out and settling to the midplane. The idea of surface accretion dates back at least to the layered accretion picture of \citet{gammie96} in which only the surface region of the disk is ionized enough to couple to magnetic fields, lose angular momentum, and thereby accrete toward the star. Surface accretion flows at supersonic velocities are predicted in recent MHD simulations of magnetized disks \citep[e.g.][]{bai13,bai17,zhu18} and may increase the temperature contrast between the disk surface and midplane \citep{mori19}, which could also have an effect on the observed gas emission. Although such gas flows in disks may be difficult to observe, observational evidence for supersonic surface accretion has been reported in at least one system: an edge on disk system, where the favorable inclination renders the surface flow more readily observable \citep{najita21}. The reason why compact disks may have enhanced surface accretion in comparison to large disks, to potentially explain the excess cool emission found in this work, is however currently not clear.

\paragraph{Other scenarios:} A number of other scenarios could be considered as possibly playing a role, including: i) if inner disk turbulence or diffusion is, for some reason, enhanced in compact disks, it might produce a larger cool water layer by mixing cool water higher up in the disk surface, or ii) planetesimal formation could perhaps trap icy solids from drifting inside the snowline region \citep{najita13}, although modeling work suggests that this effect should be minor in comparison to pebble trapping by gaps \citep{kalyaan23}. The discovery of the cool water excess with MIRI opens up interesting prospects for testing and validating these different scenarios with future modeling work.

\section{Summary \& Future prospects} \label{sec: concl}
This work reports molecular spectra observed with MIRI-MRS in four protoplanetary disks in Taurus, two compact disks (10--20~au) and two large disks (100--150~au) with multiple dust gaps observed with ALMA. The comparative analysis of these spectra, using the two large-structured disks as templates for water emission, reveals that the compact disks have excess water emission in the lower-energy lines. We summarize our conclusions as follows:

\begin{itemize}
    \item The excitation of higher-energy ($E_u > 4000$~K) rotational water lines in all four disks is very similar after accounting for the dependence of the emitting area on accretion luminosity. This hotter, inner reservoir ($T \approx 800$~K and $R_{\rm{eq}}$~$\approx 0.5$~au) is optically thick ($N \approx 1 \times 10^{18}$~cm$^{-2}$) and common across the sample, and is predicted by static thermo-chemical models that assume water formation through gas-phase reactions.
    \item The excess emission in lower-energy water lines ($E_u < 4000$~K) in the compact disks is well approximated by an LTE model with $T \approx$~170--400~K and $R_{\rm{eq}}$~$\approx 1$--10~au, therefore most likely tracing a disk region very close to ice sublimation at the water snowline. This excess emission was likely driving the correlations discovered in previous work using spectrally-blended Spitzer-IRS data \citep{najita13,banz20}.
    \item The excess cool water emission naturally fits into fundamental predictions of inner disk water enrichment through sublimation of drifting icy pebbles that cross the snowline \citep{cyr98,cc06,kalyaan21,kalyaan23}. This scenario predicts that compact disks do not have deep gaps outside the snowline to prevent pebble drift from enriching the inner disk with abundant water vapor, whereas large disks retain a large fraction of pebbles in the outer disk and decrease the inward delivery of ice. Other processes that may regulate the cool water layer to explore in the future include: a varying gas-to-dust ratio (in time and/or disk layer), the removal of water by disk gaps near the snowline, or different surface gas accretion in different disks. 
\end{itemize}

The findings of this work open up a number of exciting prospects for future work.
While this work includes the first four spectra from program GO-1640, which was set up with the specific goal of studying water emission in connection to pebble drift, a large number of disk spectra will be observed with MIRI in Cycle 1 and future cycles. The larger sample will soon enable investigations of how common the cool water excess is, how it varies with disk size and the location and depth of dust gaps, and how it may vary with other parameters like stellar mass and luminosity, age and environment, or the formation of inner disk cavities. In particular, the location and depth of the innermost disk gaps -- which should be studied in larger samples with the highest-resolution interferometric images -- may be fundamental in regulating inner disk water enrichment \citep[as proposed in][]{kalyaan23}, and may reflect in variations in the temperature, density, and radial location of the cool water excess. 

Another exciting prospect is the investigation of how molecular chemistry more generally and the elemental C/H, O/H, N/H ratios are affected by pebble drift and inner disk water enrichment, using the organics detected in disks (e.g. Figure \ref{fig: hot_components}). Future MIRI spectra will enable tests of a number of models that provide predictions on how molecular chemistry may be affected across snowlines depending on the drift and trapping or sublimation of icy particles \citep[e.g.][]{pinilla17,booth19,muller21,notsu22,cevallos22}, helping to increase our understanding of inner disk regions that are considered fundamental for planet formation \citep[e.g.][]{drazkowska23,krijt23}. 
If compact disks are confirmed to have efficient pebble drift in their inner regions, the infrared water spectrum could be used in the future as a tracer of the pebble mass flux that enters the snowline region, which will inform models on the solid mass available for the formation of planetesimals, terrestrial planets, and giant planet cores \citep[e.g.][]{lambrechts19,bitsch19_giant}. 
Future high-resolution spectroscopy at mid- to far-infrared wavelengths, beyond the MIRI-MRS cutoff at 28\,$\mu$m, will be required to more comprehensively study the cool water component down to 150\,K and possibly cooler temperatures across and beyond the snowline region \citep{notsu16,pont18,kamp21}.

\acknowledgments
This work is based on observations made with the NASA/ ESA/CSA James Webb Space Telescope. The JWST data used in this paper can be found in MAST: \dataset[10.17909/rjt1-jd72]{http://dx.doi.org/10.17909/rjt1-jd72}.
The data were obtained from the Mikulski Archive for Space Telescopes at the Space Telescope Science Institute, which is operated by the Association of Universities for Research in Astronomy, Inc., under NASA contract NAS 5-03127 for JWST. The observations are associated with JWST GO Cycle 1 programs 1549 and 1640.
This work has made use of data from the European Space Agency (ESA) mission {\it Gaia} (\url{https://www.cosmos.esa.int/gaia}), processed by the {\it Gaia}
Data Processing and Analysis Consortium (DPAC, \url{https://www.cosmos.esa.int/web/gaia/dpac/consortium}). Funding for the DPAC has been provided by national institutions, in particular the institutions participating in the {\it Gaia} Multilateral Agreement. A portion of this work was carried out at the Jet Propulsion Laboratory, California Institute of Technology, under a contract with the National Aeronautics and Space Administration.
This paper makes use of the following ALMA data: ADS/JAO.2016.1.01164.S. ALMA is a partnership of ESO (representing its member states), NSF (USA) and NINS (Japan), together with NRC (Canada), MOST and ASIAA (Taiwan), and KASI (Republic of Korea), in cooperation with the Republic of Chile. The Joint ALMA Observatory is operated by ESO, AUI/NRAO and NAOJ.

A.B. and A.K. acknowledge support from NASA/Space Telescope Science Institute grant JWST-GO-01640.
I.P. acknowledges partial support by NASA under Agreement No. 80NSSC21K0593 for the program ``Alien Earths.”
Support for F.L. was provided by NASA through the NASA Hubble Fellowship grant \#HST-HF2-51512.001-A awarded by the Space Telescope Science Institute, which is operated by the Association of Universities for Research in Astronomy, Incorporated, under NASA contract NAS5-26555.
G.R. is funded by the European Union (ERC Starting Grant DiscEvol, project number 101039651). Views and opinions expressed are however those of the author(s) only and do not necessarily reflect those of the European Union or the European Research Council Executive Agency. Neither the European Union nor the granting authority can be held responsible for them.
The authors thank the anonymous reviewer for suggestions that helped improve this manuscript. 
A.B. thanks Tommy for helping him to re-discover cold water.

\software{
Matplotlib \citep{matplotlib}, NumPy \citep{numpy}, SciPy \citep{scipy}, Seaborn \citep{seaborn}.
}

\appendix

\begin{figure*}
\centering
\includegraphics[width=1\textwidth]{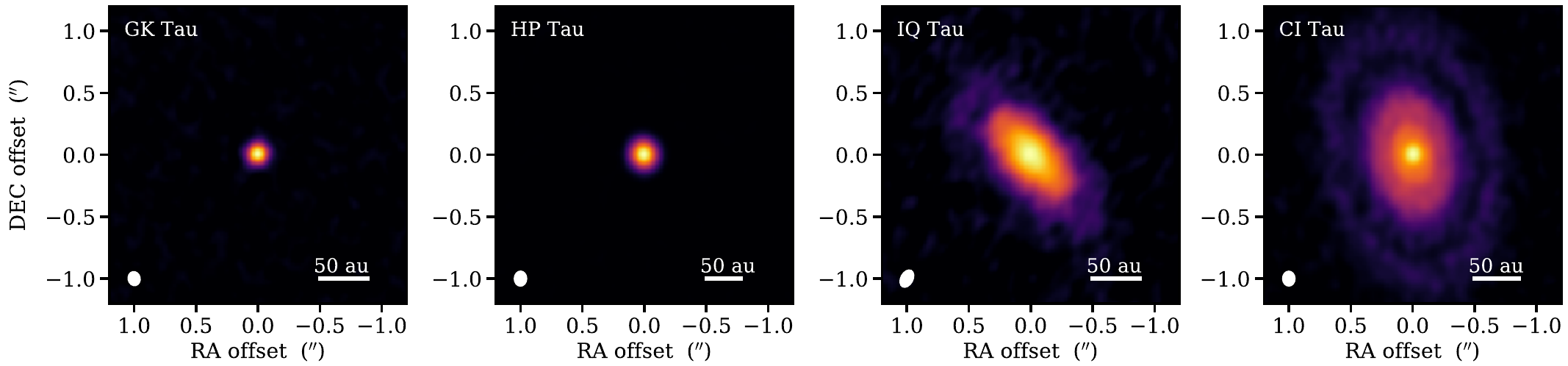} 
\caption{ALMA images for the sample ordered by disk radius, from \citet{long19}.
}
\label{fig: ALMA}
\end{figure*}

\begin{figure*}
\centering
\includegraphics[width=1\textwidth]{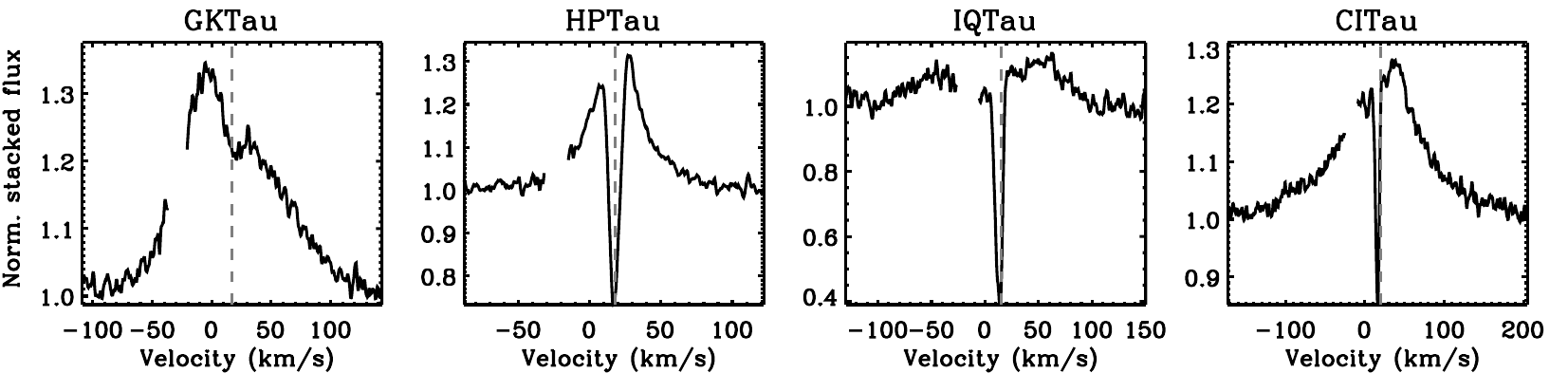} 
\caption{Stacked P5--P18 line profiles of CO ro-vibrational fundamental emission observed in the four targets, from iSHELL and CRIRES spectra in \citet{banz22, brown12} and this work. The stellar RV is marked with a dashed line, and the gaps on the blue side of each line are due to telluric lines at the time of observation.
}
\label{fig: CO_lines}
\end{figure*}

\section{Ground-based data for the sample} \label{app: ground data}
Figure \ref{fig: CO_lines} reports high-resolution (R~$\sim$~60,000--90,000, or 3--5~km/s) line profiles for the CO fundamental band ($\nu = 1-0$) as observed from the ground with iSHELL \citep{ishell22} as part of an ongoing disk survey \citep[][with GK~Tau and HP~Tau observed on February 15, 2023, just 12 days before the MIRI spectra analyzed in this work]{banz22} and CRIRES \citep{crires} as part of a previous survey \citep[IQ~Tau as observed in 2008, from][]{pont11,brown12}. The CO kinematics, which trace a similar range of inner disk regions as the infrared water lines \citep{banz23}, demonstrate that CO spans similar disk radii in these four disks around 0.1~au (in Table \ref{tab: sample} we report the Keplerian radii from the measured velocities at 10\% and 50\% of the CO line width). Some notable differences are visible in IQ~Tau (showing more compact emission) and HP~Tau (showing more extended emission). All four disks show narrow CO absorption at low velocities that might be due to self-absorption in a cooler molecular inner disk wind above the disk surface \citep[][the absorption is detected only up to P4 in GK~Tau, and therefore not visible in Figure \ref{fig: CO_lines}]{pont11,banz22}.

\begin{figure*}
\centering
\includegraphics[width=1\textwidth]{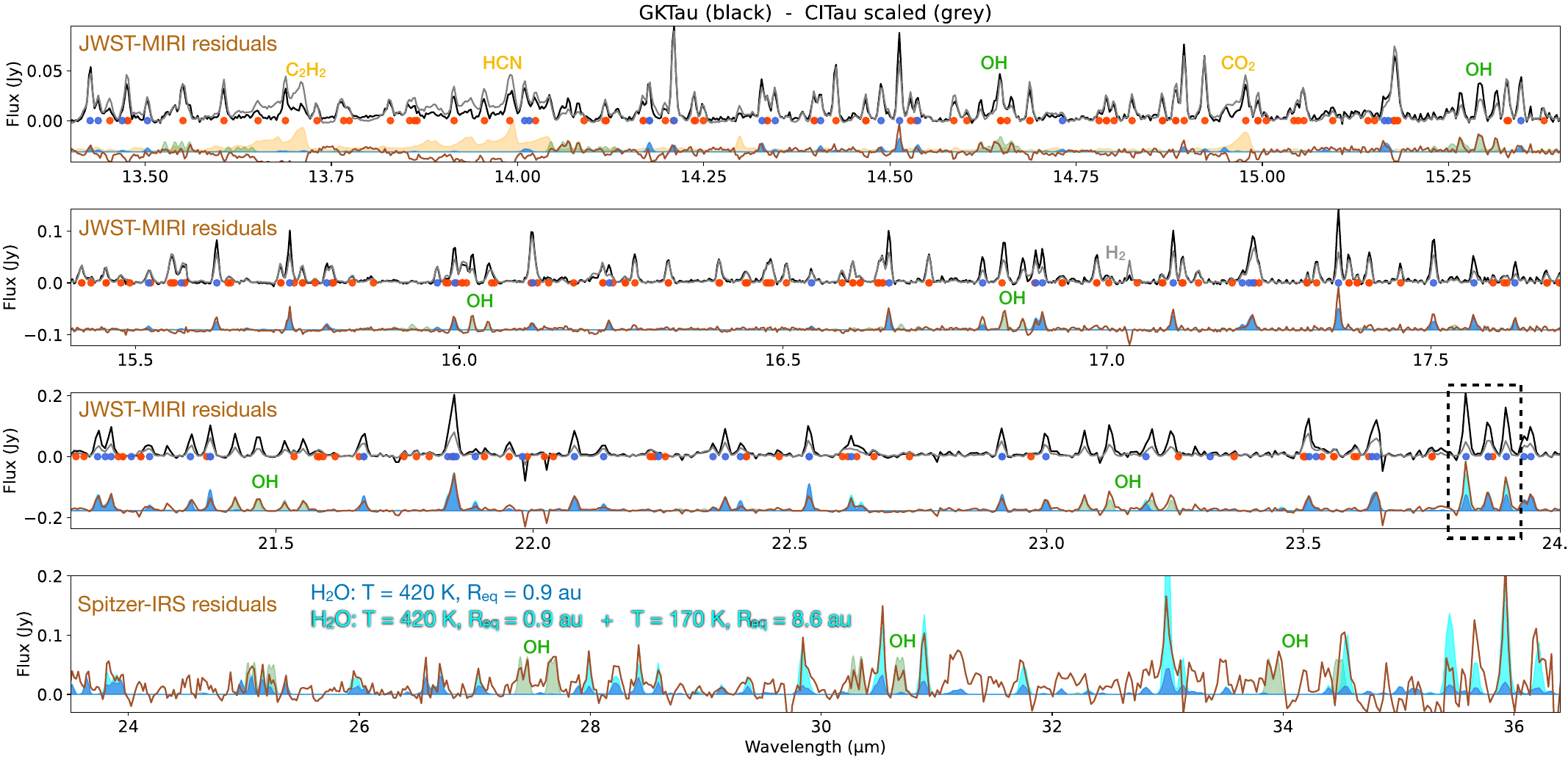} 
\caption{Broader view of excess cool component in rotational water emission in GK Tau, using CI Tau as template for the hot component as in Figure \ref{fig: subtraction_zoomed}. The same scaling and subtraction procedure described in Section \ref{sec: analysis} is applied here, where residuals after subtraction are shown in brown for MIRI (top 3 panels) and IRS (bottom panel). A two temperature fit is shown in blue (only the $T \approx 400$~K component) and cyan ($T \approx 400$~K and $T \approx 170$~K components together), with best-fit values as reported in Appendix \ref{app: cool component}. The dashed box marks the two MIRI lines discussed in the text. Other molecules are reported and labelled in different colors for reference.}
\label{fig: residual_spitzer}
\end{figure*}

\section{Broader spectral characterization of excess cool component} \label{app: cool component}
To obtain a more comprehensive view of the cool excess component in compact disks, we include the Spitzer-IRS spectra of CI~Tau and GK~Tau from \citet{pont10,banz20}. HP~Tau was never observed with IRS, and IQ~Tau only has a lower quality spectrum with higher noise \citep{james22}.
As shown in \citet{kzhang13,blevins16}, cool water emission from the snowline at $T \approx 150$~K should dominate the observed water spectrum at 30--70~$\mu$m, beyond the MIRI range but partly covered by IRS (see Figure \ref{fig: water_radial_props}). We apply to the IRS spectra the same scaling and subtraction procedure developed on the MIRI spectra (Section \ref{sec: analysis}), and re-fit the excess emission in GK~Tau at MIRI and IRS wavelengths with a two temperature model to approximately account for a radial temperature gradient that is expected at the disk radii across the snowline \citep[1--10~au,][]{blevins16}. The fit is shown in Figure \ref{fig: residual_spitzer}; the two components have: $T \approx 420$~K, $N \approx 3.5 \times 10^{17}$~cm$^{-2}$, and $R_{\rm{eq}}$~$\approx 0.9$~au (best-fit values that are essentially unchanged from fitting the excess in the MIRI spectrum alone, as reported in Section \ref{sec: analysis}), and $T \approx 170$~K, $N \approx 3 \times 10^{16}$~cm$^{-2}$, and $R_{\rm{eq}}$~$\approx 9$~au. The colder of these two components emits most prominently at $>30$~$\mu$m as found before \citep{kzhang13,blevins16}, but the IRS spectra alone would leave large uncertainties on its presence and properties.
The new MIRI data analyzed in this work, instead, allow to identify the 170~K component by de-blending individual lines (just like it was shown in Figure \ref{fig: subtraction_zoomed} for the separation of the $T \approx 800$~K and $T \approx 400$~K components): two prominent MIRI lines at 23.81~$\mu$m and 23.89~$\mu$m clearly show excess that could not be fitted with the rest of the excess spectrum using a single temperature of $T \approx 400$~K, and their excess flux is now filled in by the extra cool component at $T \approx 170$~K (Figure \ref{fig: residual_spitzer}). From Figure 6 in \citet{kzhang13}, we can see that this was to be expected: these two lines are as sensitive to the abundance and radial distribution of water across the snowline just like the lines at 30~$\mu$m (covered by IRS) and 60-70~$\mu$m (covered by Herschel-PACS), which indeed all share similar properties, $E_u \approx 1000$~K and $A_{ul} \approx$~1-10~s$^{-1}$.

\section{Single-temperature fits to the high-energy lines} \label{app: hot component}
\begin{figure*}
\centering
\includegraphics[width=0.85\textwidth]{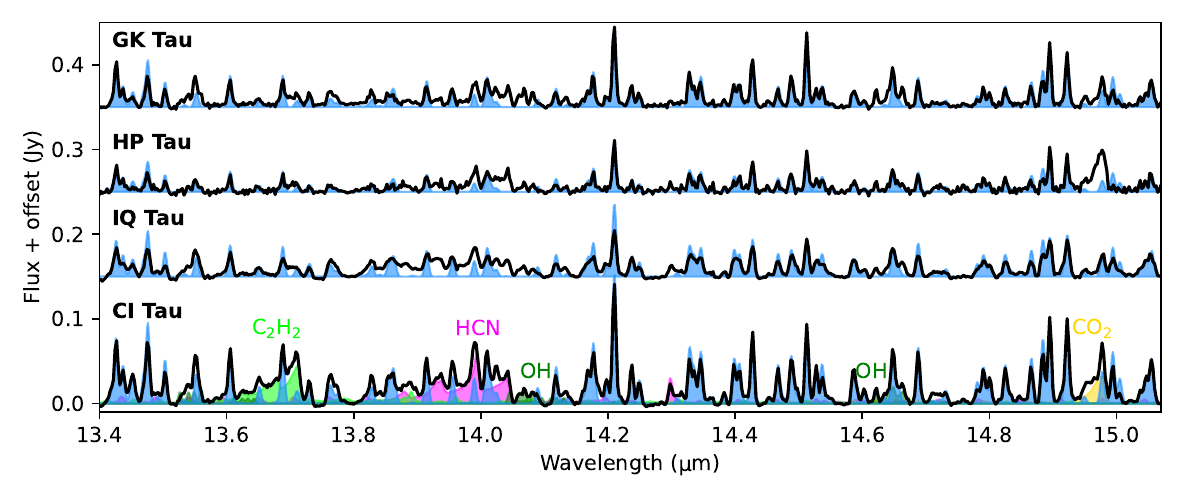} 
\caption{Single-temperature LTE fits to the high-energy lines that dominate water spectra at 12--16~$\mu$m. Best-fit parameters are reported in Table \ref{tab: slab_results}. Representative models of organics and OH emission are reported for reference at the bottom.}
\label{fig: hot_components}
\end{figure*}

For comparison to the properties of the excess cool component, we also fit a single-temperature LTE model to the higher-energy water lines that dominate the emission at 12--16~$\mu$m. The best-fit parameters cluster around $T \approx$~830~K, $N \approx 1 \times 10^{18}$~cm$^{-2}$, and equivalent emitting radius $R_{\rm{eq}}$~$\approx 0.5$~au and are reported in Table \ref{tab: slab_results} and Figure \ref{fig: hot_components}. At this column density, the high-energy water lines at these wavelengths are moderately optically thick with an opacity at the line center $\tau \approx$~0.2--10 \citep[for the definition of $\tau$, see appendix in][]{banz12}. The best-fit emitting areas confirm what reported in Section \ref{sec: analysis} about the high-energy lines mostly tracing a luminosity-dependent optically thick emitting area in different disks, with CI~Tau having the largest emitting area with $R_{\rm{eq}}$~$\approx 0.65$~au, and the other three disks having similar emitting area with $R_{\rm{eq}}$~$\approx 0.4$~au. The $L_{\rm{acc}}$ ratio between CI~Tau and the other three disks is $\approx$~2.5--5, which elevated to the power of 0.6 (the slope of the correlation in Figure \ref{fig: Lacc_correlations}) gives $\approx$~1.7-2.6, fully consistent with the emitting areas ratio of $\approx 2.1$ from the slab fits.

The best-fit values found for the hot component are consistent with LTE fits to spectrally-resolved water emission lines near 12.4~$\mu$m from ground-based observations \citep{banz23}, and with recent fits to a similar range of MIRI water lines in another disk \citep{grant23}. A single temperature in LTE is generally sufficient to provide a good fit in this limited wavelength range as found before with Spitzer and MIRI spectra \citep{cn11,banz12,grant23}, but fitting larger ranges requires at least a temperature gradient \citep{blevins16,liu19,banz23}. For reference, we report that preliminary two-temperature LTE fits to the whole range of rotational lines in MIRI spectra of these disks are also consistent with the two temperatures reported in this work, a hot component with $T \approx$~800--1000~K and a cool one with $T \approx$~200--400~K.
Previous work suggested that the gas density in the water-emitting layer is below the critical density $n_{\rm{crit}}$ necessary to thermalize infrared water lines, especially at higher upper level energies of $E_u > 4000$~K \citep{meijerink09}, with rotational lines having $n_{\rm{crit}} \approx 10^{8-11}$~cm$^{-3}$ and the ro-vibrational bands having $n_{\rm{crit}} \approx 10^{12-16}$~cm$^{-3}$. This should produce non-LTE populations of the higher-energy levels which should be most visible in the ro-vibrational bands \citep{bosman22}, explaining a strong flux reduction that has indeed been observed \citep{banz23}; the rotational lines with higher $E_u$ may also be reduced by factors of a few \citep{meijerink09,banz12}. The increased resolution of MIRI now better shows residual mismatch with the LTE models, which are currently being investigated in terms of non-LTE excitation and will be reported in a future paper in combination to the analysis of the ro-vibrational band at $< 9$~$\mu$m.

\bibliography{1640_paper1}{}
\bibliographystyle{aasjournal}

\end{document}